\documentclass{optica-article}

\journal{opticajournal} 
\articletype{Research Article}

\usepackage{lineno}

\usepackage{siunitx}
\usepackage{graphicx}
\usepackage{dcolumn}
\usepackage{bm}
\usepackage{xcolor}
\usepackage{soul}
\usepackage{amsmath}
\usepackage{mathrsfs}
\usepackage{amsfonts}
\usepackage{upgreek}
\usepackage{url}
\usepackage{appendix}

\begin{document}

\title{Single-shot in situ pulse-duration measurement using plasma grating}

\author{Jimin Wang,\authormark{1} Yanlei Zuo,\authormark{1} Kainan Zhou,\authormark{1} Zhaoli Li,\authormark{1}
Pengyu Wei,\authormark{1,2} Xiao Wang,\authormark{1} Jie Mu,\authormark{1} Xiaodong Wang,\authormark{1}
Xiaoming Zeng,\authormark{1} Zhaohui Wu,\authormark{1,6} Hao Peng,\authormark{3,7} C.\ Riconda,\authormark{4}
and S.\ Weber\authormark{5}}

\address{\authormark{1}National Key Laboratory of Plasma Physics, Laser Fusion Research Center, China Academy of Engineering Physics, Mianyang 621900, China\\
\authormark{2}University of Science and Technology of China, Hefei 230026, China\\
\authormark{3}Shenzhen Key Laboratory of Ultraintense Laser and Advanced Material Technology, Center for Advanced Material Diagnostic Technology,
and College of Engineering Physics, Shenzhen Technology University, Shenzhen 518118, China\\
\authormark{4}LULI, Sorbonne Universit{\'e}, CNRS, Ecole Polytechnique, CEA, 75005, Paris, France\\
\authormark{5}Extreme Light Infrastructure ERIC, ELI-Beamlines Facility, 25241 Doln{\'\i} B{\u r}e{\u z}any, Czech Republic\\
}
\email{\authormark{6}wuzhaohui20050@163.com}
\email{\authormark{7}penghao@sztu.edu.cn}

\begin{abstract*}
Accurate measurement of the pulse duration of ultrashort, ultra-intense laser pulses at focus is essential for strong-field science. Most existing diagnostics, however, cannot allow direct in situ measurement in the focal region because of damage-threshold limits and unavoidable spatial averaging. We present a direct single-shot far-field diagnostic based on a plasma grating. In this method, the pulse duration is encoded in the axial length of an interference-written plasma grating and retrieved from the corresponding Bragg-diffraction signal. Comparison with near-field (pre-focus) autocorrelator measurements and far-field (at-focus) scanning measurements confirms single-shot pulse-duration retrieval in the focal region over 35--130 fs, and the method remains effective at a peak intensity of $\sim 10^{16}{\rm W/cm^2}$. In principle, the measurable range can be extended to 15--300 fs and to higher peak intensities. The method is insensitive to the laser central wavelength and offers a practical approach to far-field diagnostics in high-power laser systems.
\end{abstract*}

\section{Introduction}

With ongoing advances in ultrashort, ultra-intense laser technology, pulse durations have been pushed to the femtosecond and attosecond regimes, while peak intensities have reached the relativistic level of $10^{23}{\rm W/cm^2}$ \cite{Li2025}. These capabilities have made lasers powerful tools for studying fundamental physical processes. They are now widely used in areas such as relativistic plasma physics \cite{Palaniyappan2012}, laboratory astrophysics \cite{Chen2015}, laser-driven particle acceleration \cite{Macchi2013}, and high-order harmonic generation \cite{Teubner2009,Thaury2010}. Lasers are also central to studies of ultrafast dynamics in atomic and molecular physics, life sciences, and chemical reactions \cite{Zewail1988}.

In studies of laser--matter interaction, accurate knowledge of the pulse duration is needed for reliable estimation of the peak intensity in the focal region and for further exploration of the underlying physics. Common pulse-duration diagnostics include intensity autocorrelation, frequency-resolved optical gating (FROG) \cite{Trebino1997,Shea2001,Bates2010}, spectral phase interferometry for direct electric-field reconstruction (SPIDER) \cite{Radunsky2007}, self-referenced spectral interferometry (SRSI) \cite{Trisorio2012,Miranda2012,Oksenhendler2010,Iaconis1998}, and dispersion-scan (D-scan) \cite{Loriot2013}. In ultra-high-power laser systems based on chirped-pulse amplification (CPA), however, measuring the pulse duration directly in the focal region remains difficult.

A common workaround is to sample the beam before focus, recollimate it, and measure the pulse duration with one of the methods above. The pulse properties at focus are then inferred through a transfer function. In systems that use large-aperture optics \cite{Strickland1985}, however, non-ideal beam quality, optical aberrations, and material dispersion can introduce substantial uncertainty into this transfer function, leading to errors greater than 50\% \cite{Long2022}.

Direct measurement in the focal region could in principle avoid these problems, but several basic limitations have so far prevented routine implementation. These techniques rely on nonlinear processes that can easily saturate at extremely high power density and may irreversibly damage nonlinear crystals. Under practical experimental conditions, it is also difficult to achieve precise spatial overlap between the focused spot and the nonlinear crystal.

For high-energy laser systems, which typically operate at low repetition rate and still exhibit shot-to-shot fluctuations, a diagnostic capable of direct single-shot pulse-duration measurement in the far field is therefore needed.

Early methods used two-photon fluorescence \cite{Giordmaine1967}, where excitation of a nonlinear medium such as a dye or fused silica produced a fluorescence pattern proportional to the pulse autocorrelation. In practice, however, background fluorescence from photographic film severely reduced the signal-to-noise ratio (SNR), making accurate measurement and real-time monitoring difficult \cite{Bradley1974}. Fischer et al. later demonstrated a single-shot autocorrelation method based on transverse second-harmonic generation in strontium barium niobate (SBN) crystals \cite{Horowitz1993,Fischer2007,Saltiel2008}. With optimized imaging optics, including spherical lenses, this method enabled background-free single-shot far-field autocorrelation measurements of complex pulse structures \cite{Yu2018}.

Even so, crystal-based methods are still limited by the damage threshold of the crystal and by intrinsic wavelength sensitivity over the $0.4$--$6\mu\mathrm{m}$ band. These limitations make them difficult to use for focal-region pulse-duration characterization in mainstream solid-state lasers operating near $800\mathrm{nm}$, where the peak focal intensity commonly exceeds the tolerance of solid materials.

To avoid the limitations of solid media, Noam Shlomo and Eugene Frumker recently proposed an in situ pulse diagnostic based on the conductivity of laser-ionized gas, combined with an ionization model to retrieve the pulse intensity and duration \cite{Shlomo2025}. This method, however, requires a relatively complex vacuum system and accurate ionization modeling, which increases experimental complexity and introduces additional uncertainty in practical use.

Against this background, plasma gratings offer a promising route for characterizing high-intensity laser pulses in the focal region. The damage threshold of laser-generated plasma is about six orders of magnitude higher than that of conventional optical components \cite{Malkin1999}. On this basis, plasma mirrors \cite{Thaury2007} and plasma gratings \cite{Edwards2023} have shown strong reliability and have been applied in a range of areas, including spatial intensity modulation \cite{Leblanc2017,Edwards2024a}, pulse compression \cite{Wu2022,Li2023,Wu2024}, dispersion control \cite{Edwards2022}, and optical switching \cite{Edwards2024b}.

One of the main mechanisms for plasma-grating formation is spatially varying ionization (SVI) \cite{Suntsov2009,Shi2011,Durand2012,Zhang2021}, in which the peak intensity of the interference fringes exceeds the ionization threshold of the medium. This mechanism makes it possible to map pulse duration from the time domain into the spatial domain, because the spatial extent of the interference region is directly related to the pulse duration. Previous work has shown that the SVI scheme requires only millijoule-level femtosecond pump pulses to generate plasma gratings through ionization-rate differences within standing-wave interference patterns \cite{Edwards2022}. Experiments have reported SVI plasma gratings with densities up to $2 \times 10^{19}{\rm cm^{-3}}$ \cite{Shi2011} and diffraction efficiencies as high as 36\% \cite{Edwards2023,Waczynski2024}. The plasma lifetime is on the order of tens to hundreds of picoseconds \cite{Durand2012}. During the early stage, from the onset of ionization to the first several tens of picoseconds, recombination and diffusion are negligible, so the plasma density can be treated as effectively constant. This behavior meets the requirements of the present measurement scheme.

In the method proposed here, pulse-duration information is encoded in the spatial extent of the plasma grating and then read out with a probe beam. The approach enables accurate single-shot characterization of laser pulses with durations from tens to hundreds of femtoseconds. It avoids the damage-threshold limits of conventional optics, suppresses background noise, and removes wavelength-dependent restrictions. As a result, it provides a precise and high-SNR route for temporal diagnostics in the optimization and operation of ultra-high-peak-power laser systems.

\section{Scheme and experimental setup}

\subsection{\label{scheme}Scheme}

\begin{figure}
    \centering
    \includegraphics[width=10cm]{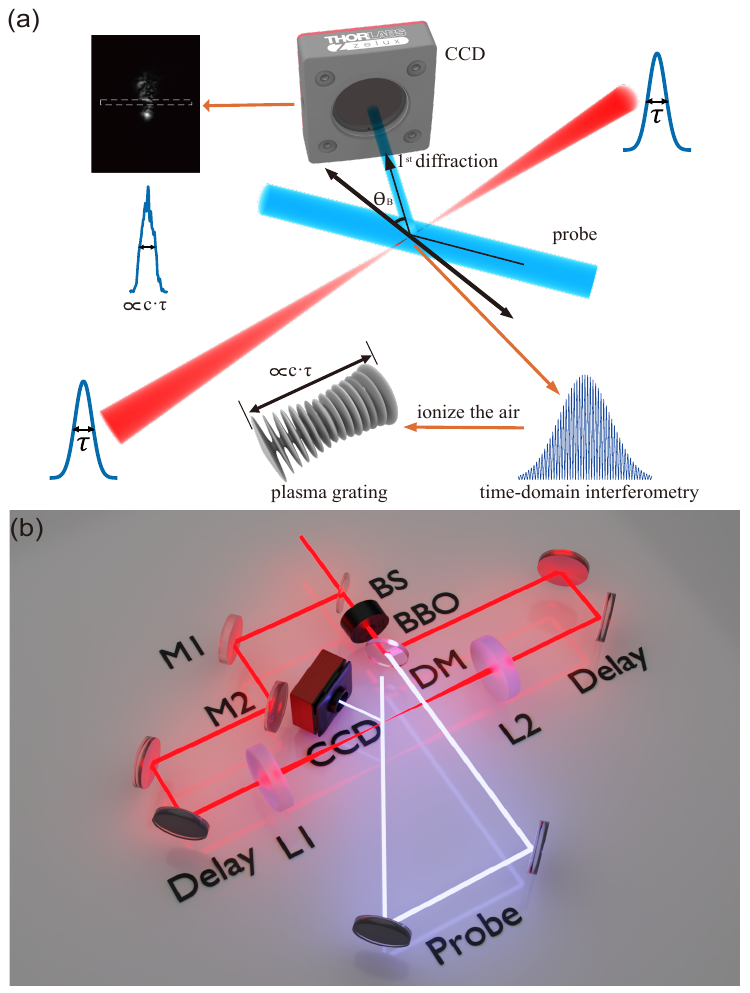}
    \caption{(a) Schematic diagram of the principle for measuring far-field laser pulse duration using holographic plasma grating. (b) Schematic of experimental setup.}
    \label{fig:1}
\end{figure}
Hereafter, the measured pulse is referred to as the \textit{signal pump}, the other pulse involved in interference-driven ionization as the \textit{reference pump}, and the readout pulse as the \textit{probe}. Consider two pump pulses with central wavelength $\lambda_p$ intersecting at an angle of $2\theta_p$. In the overlap region, the interference produces a spatial intensity modulation with a period of $\Lambda = \frac{\lambda_p}{2\sin(\theta_p)}$ \cite{Suntsov2009}. The SVI plasma grating serves as a temporal encoder for retrieving the pulse duration in the focal region. This principle does not depend on the central wavelength of the signal pump, whereas the probe only needs to satisfy the Bragg condition \cite{Shi2011}. The method is therefore applicable to ultrashort-pulse laser systems over a broad range of central wavelengths.

The measurement principle is illustrated in Fig.\ref{fig:1}(a). The test (pump) pulse is split into two time-synchronized, counter-propagating beams that interfere at the focus to ionize the gas and form a plasma transmission grating. A narrowband probe illuminates the grating at the Bragg angle $\theta_B = \sin^{-1}\left(\lambda_{\mathrm{pr}}/(2n_{\mathrm{med}}\Lambda)\right)$, where $\lambda_{\mathrm{pr}}$ is the probe wavelength and $n_{\mathrm{med}}$ is the refractive index of the gas medium at the probe wavelength. The first-order diffracted beam is then imaged onto a CCD by an imaging system. Since diffraction occurs only where the plasma grating is present, the CCD image records both the grating extent and the diffraction efficiency. From the two-dimensional diffraction-intensity distribution, the axial envelope of the $+1$ diffraction order is obtained, and its full width at half maximum is used to define the effective grating length $L$.

To measure the pulse duration accurately, we examine how the temporal envelope of the pump pulse is encoded in the plasma grating and how the grating length can be determined from the $+1$-order diffraction intensity. A quantitative relationship among laser intensity, pulse duration, and grating length is then established, and additional factors that may affect the diffraction efficiency of the grating are taken into account.

\subsubsection{Theoretical model}

Two ultrashort pulses (central wavelength $\lambda_p$) propagate in opposite directions ($\theta_p = 90^\circ$) along the z-axis in a gaseous medium. When focused, the pulses overlap in the focal region, where their electric fields superpose and form a standing-wave interference pattern. Because the spatiotemporal overlap region is much shorter than the Rayleigh range of a Gaussian beam, it is assumed that the transverse spatial profile of the pulses does not change appreciably in the focal region and that the pulse-width information is essentially preserved. Under these assumptions, the problem can be reduced to a one-dimensional model.

If the instantaneous intensity of each pulse is $I_0(t)$, the total instantaneous intensity in the overlap region is
\begin{equation}
  \begin{aligned}
    I(z,t) &= I_0(t) + I_0(t) + 2\sqrt{I_0^2(t)}\cos(2kz)= 2I_0(t)\bigl[1 + \cos(2\pi z/\Lambda)\bigr],
  \end{aligned}
\end{equation}
where $k=2\pi/\lambda_p,\ \ \Lambda=\pi/k=\lambda_p/2$ is the spatial interference period.In the counter-propagating geometry, the longitudinal coordinate $z$ along the grating axis is in one-to-one correspondence with the relative delay between the two pump pulses. Writing the two pump envelopes as $E_{+}(t-z/c)$ and $E_{-}(t+z/c)$, the local temporal mismatch at position $z$ is
\begin{equation}
  \Delta t(z)=\frac{2z}{c},
  \label{eq:dtz}
\end{equation}
so that $z=0$ corresponds to perfect temporal overlap, while increasing $|z|$ corresponds to a larger pump--pump delay. The intensity is therefore highest near the spatiotemporal overlap center ($z\approx0$) and decreases away from the center as the pulse envelope decays.

The time evolution of the free-electron density in the focal region can be described by a rate equation:
\begin{equation}
  \frac{dN_e(z,t)}{dt} = W\big(I(z,t)\big)[N_0 - N_e(z,t)],
\end{equation}
where $N_0$ is the neutral particle number density and $W(I)$ is the ionization rate (the MO-PPT model is used later in this work). Because of the periodic spatial modulation, the ionization-produced electron density can be expanded in a spatial Fourier series,
$N_e\left(z,t\right)=N_{e,0}\left(t\right)+N_{e,1}\left(t\right)\cos{\left(\Delta kz+\phi\right)}+N_{e,2}\left(t\right)\cos{\left(2\Delta kz+\cdots\right)}+\cdots$,
where the first-order electron-density grating component is the key quantity for the subsequent Bragg-diffraction measurement of pulse duration. After the signal pump and reference pump pass through the focal region, the axial electron-density distribution in the overlap region can be regarded as the time integral of the ionization response: $N_{e,1}\left(z\right)\sim\int_{-\infty}^{+\infty}{\mathcal{F}_1[W(I(z,t))]} dt$.

The plasma induces a refractive-index change, $\mathrm{\ }\Delta n\simeq-\frac{\Delta N_e}{2N_{\mathrm{cr}}\left(\lambda\right)}$, where $N_{\mathrm{cr}}\left(\lambda\right)=\frac{\varepsilon_0m_e\left(2\pi c/\lambda\right)^2}{e^2}$ is the critical plasma density at the probe frequency. Therefore, the first-order component of the refractive-index modulation $\Delta n_1\left(z\right)$ is proportional to $N_{e,1}(z)$.

When a probe beam is incident on the plasma grating at the Bragg angle, the first-order diffraction efficiency under Bragg-matched two-wave coupling for a thick grating can be expressed as \cite{yeh1994introduction,boyd2008nonlinear}
\begin{equation}
  \eta_{+1}=\frac{I_{+1}}{I_{\mathrm{pr}}}\approx \sin^2(\kappa L_g),\
  \kappa=\frac{\pi\Delta n_1}{\lambda_{\mathrm{pr}}\cos\theta_B},
  \label{eq:cwt_eta}
\end{equation}
where $I_{\mathrm{pr}}$ is the incident probe intensity, $\lambda_{\mathrm{pr}}$ is the probe wavelength, and $L_g$ is the effective interaction length, determined by the grating thickness and the readout overlap length.

In the weak-grating, small-signal limit $\kappa L_g \ll 1$, one obtains
\begin{equation}
  \eta_{+1}\approx (\kappa L_g)^2 \propto \left|\Delta n_1\right|^2 \propto \left|N_{e,1}\right|^2,
  \label{eq:eta_small_signal}
\end{equation}
so that, provided the probe intensity is far below the ionization threshold of the gaseous medium and the imaging system has sufficient axial resolution, the measured axial envelope $I_{+1}(z)$ can be used to represent the spatial distribution of $\left|N_{e,1}(z)\right|^2$ and thereby define the grating length $L$. In this work, $L$ is taken as the full width at half maximum of $I_{+1}(z)$.

\begin{figure}[!htbp]
    \centering
    \includegraphics[width=10cm]{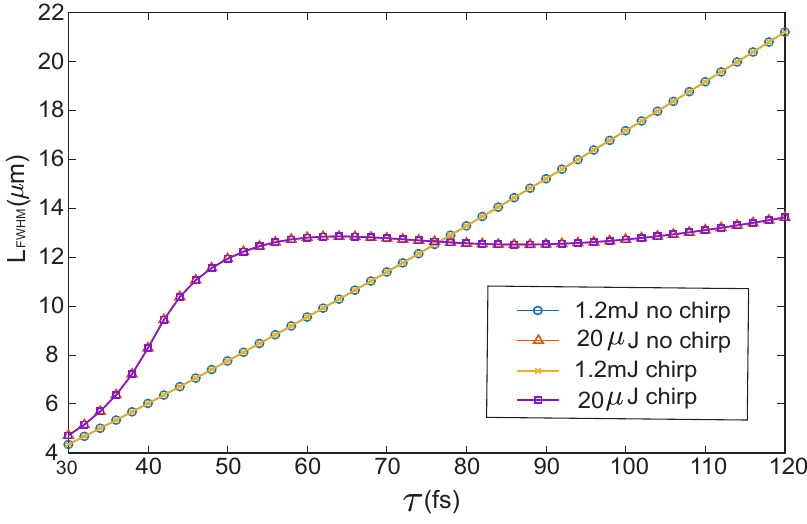}
    \caption{Calibration curves between the pulse duration of the signal pump and the resulting plasma-grating length; different colors and symbols correspond to different pulse energies. 'Chirp' indicates that the pulse duration is varied by adding dispersion, whereas 'no chirp' indicates that the Fourier-transform-limited pulse duration is varied.}
    \label{fig:2}
\end{figure}

\subsubsection{Calibration and readout}

As described above, the effective grating length is defined as $L \equiv \mathrm{FWHM}\left[I_{+1}(z)\right]$. In experiments, the directly measured quantity is the axial envelope of the $+1$ diffraction-order intensity $I_{+1}(z)$. Because it is difficult to derive a closed-form analytical relation between $L$ and the pulse duration $\tau$, we construct a numerical calibration curve $L(\tau)$ and determine the pulse duration by inverting the measured value of $L$. In chirped-pulse amplification (CPA) laser systems, the pulse duration is tuned by adjusting the spectral dispersion or the effective bandwidth. Figure \ref{fig:2} shows typical calibration curves for chirped pulses and Fourier-transform-limited pulses at different pulse energies. At the microjoule ($\mu\mathrm{J}$) level, the response is approximately linear only over 30--50 fs and saturates outside this interval. In contrast, when the pulse energy reaches the millijoule (mJ) level, the curves exhibit an approximately linear correlation over 30--120 fs.

The shape of the calibration curve is governed by two competing effects. For a fixed pump energy and focusing geometry, increasing the pulse duration extends the interaction time while reducing the peak intensity. Since the ionization rate $W(I)$ is highly nonlinear, shorter pulses usually generate higher peak intensities and a wider effective ionization region. As $\tau$ increases, the effective writing range of $N_{e,1}(z)$ broadens, and $L(\tau)$ increases monotonically, with an approximately linear region under suitable conditions. With a further increase in pulse duration, however, the lower peak intensity causes the local intensity at some axial positions to fall below the effective ionization threshold, which slows the expansion of the written region. As a consequence, the increase in $L(\tau)$ becomes weaker and gradually approaches saturation, as seen in the microjoule-level curves in Fig.\ \ref{fig:2}. By choosing an appropriate pulse energy and gas medium, a one-to-one relation between grating length and pulse duration can therefore be established, allowing the pulse duration to be determined from the measured grating length using the inversion procedure described above. The ionization model used to generate the calibration curves, together with the numerical simulation workflow and the retrieval procedure, is provided in Appendix A--C.

To ensure accurate readout of the grating length, the probe conditions must be chosen carefully. In the experiment, the probe beam is incident on the preformed plasma grating at the Bragg angle, so the readout process is essentially analogous to a pump--probe measurement. This readout can be perturbed if the probe pulse is too intense or if the probe delay is not selected properly. For example, at a delay of 3.5 ps, collisional ionization and inverse Bremsstrahlung can significantly modify key plasma parameters, including the electron-density distribution \cite{Wahlstrand2011,Shi2011}, and thereby degrade the grating structure. To minimize such perturbations and preserve a reliable correspondence between the diffraction signal and grating length, we use a frequency-doubled probe with an intensity far below the ionization threshold ($<10^{12}{\rm W/cm^2}$) and limit the probe delay to 1--2 ps. Under these conditions, the probing process does not significantly alter the plasma state. Previous work has also shown that a transmissive plasma grating does not introduce appreciable distortion into the spatiotemporal properties of the laser pulse \cite{Edwards2022}.

\subsection{Experimental setup}

As shown in Fig.\ 1(b), the scheme was implemented on a Ti:sapphire CPA system operating at 10Hz. The system delivered pulses as short as $\sim 30$fs, with energies up to 12mJ at a central wavelength of 800nm, and exhibited a super-Gaussian-like spectrum with an 80 nm FWHM.

The input beam first passed through an iris aperture to control the beam diameter. After beam reduction, the beam was coupled into the measurement optical path (omitted in the optical layout). A 1:1 beam splitter (BS) divided the beam into two arms. In the first arm, the beam served as the signal pump. It was reflected by mirrors M1 and M2, passed through a delay line for fine timing adjustment, and was then focused into ambient air by lens L1 ($f/10$). In the second arm, a BBO crystal generated a second-harmonic beam at 400nm with a measured conversion efficiency of approximately 1\%. Numerical simulations show that a 1\% energy difference between the reference pump and the signal pump has a negligible effect on both the grating length and the diffraction efficiency, as detailed in Appendix D. The two pump energies were therefore treated as equal under the experimental conditions. A dichroic mirror (DM) then separated the 800nm fundamental beam (reference pump) from the 400nm probe beam. At the focus, the counter-propagating 800nm signal pump and reference pump interfered and ionized the gas, forming a volume plasma transmission grating. The 400nm probe illuminated the grating near the Bragg angle ($\sim 30^\circ$).

After second-harmonic generation and spectral filtering, with the DM removing the residual fundamental light, the 400 nm probe was effectively narrowband ($\sim 5$ nm). This narrow bandwidth prevents the divergence of the diffracted beam from blurring the recorded spot. After collimation, the probe beam diameter was set to 5 mm, much larger than the several-tens-of-micrometers scale of the grating. This difference ensures approximately uniform illumination across the entire grating region.

In the detection path, the first-order diffracted beam at $\sim 30^\circ$ was imaged onto a CCD by a lens. Using the object--image relation for a single lens, together with calibration by a reference target placed at the plasma-grating plane, the object--image magnification in the present configuration was determined to be $M=7$. The magnification of the single-lens imaging system strongly affects the accuracy of pulse-duration retrieval. The pixel size of a typical CCD camera is on the order of several micrometers, whereas the grating length is only several tens of micrometers. If the magnification is too small, the integrated intensity profile cannot resolve the full width at half maximum (FWHM) with sufficient accuracy. For this reason, a magnification greater than 5 is used to provide adequate spatial resolution for reliable characterization.

The two-dimensional spot profile in the $x$ and $y$ directions provides the grating length and width, allowing quantitative retrieval of the far-field pulse duration and the one-dimensional spatial intensity distribution. Unlike pulse-measurement schemes based on nonlinear crystals, where the second-harmonic signal often overlaps with the fundamental light, the present method spatially separates the zero-order background through the incident-beam geometry. This separation suppresses background noise and improves the signal-to-noise ratio (SNR).

The pulse duration was adjusted by changing the grating separation in the compressor, while the energies of the signal pump and reference pump were kept approximately constant. To probe a stable plasma grating, a delay of about $1 \mathrm{ps}$ was introduced for the $400 \mathrm{nm}$ probe pulse. On this timescale, plasma diffusion and recombination are negligible, and no spatial modulation of the plasma density is produced by pump--probe interference \cite{Wahlstrand2011,Durand2012}.

\section{Results and discussion}

\subsection{Single-shot measurement and basic validation}

\begin{figure}
    \centering
    \includegraphics[width=10cm]{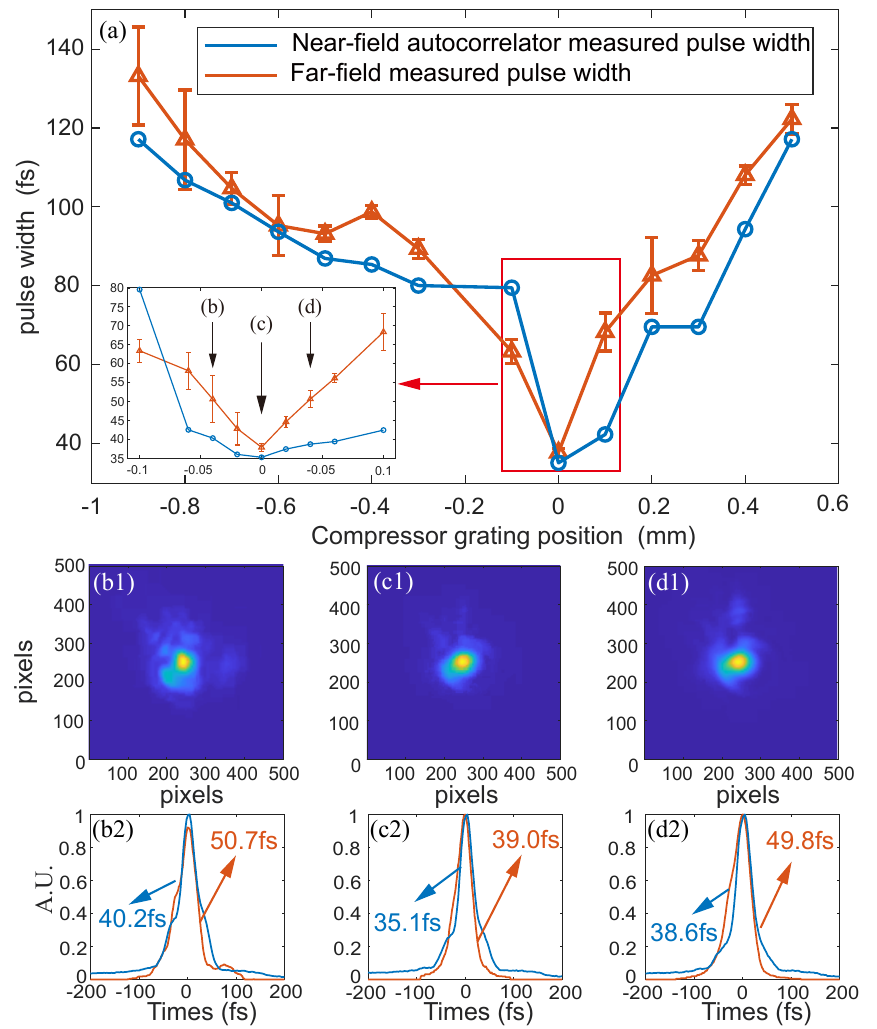}
    \caption{(a) Measured pulse duration as a function of compressor grating displacement. (b1)--(d1) Diffraction patterns recorded at compressor grating positions of $-0.04$mm, 0mm, and 0.04mm, respectively. (b2)--(d2) Corresponding temporal waveforms obtained from near-field autocorrelator measurements (blue curves) and plasma-grating diffraction analysis (orange curves) at the same three positions.}
    \label{fig:3}
\end{figure}

The experimental verification covers a pulse-duration range from 35 to 130 fs. In principle, the method can be extended to a wider pulse-duration range, although the practical limit is usually set by the Rayleigh length of the focused Gaussian beam.

The pulse duration was first measured in the near field with an autocorrelator. By coarsely adjusting the grating separation in the compressor, the setting that provided optimal chirp compensation was identified and taken as the initial grating position. The grating separation was then scanned on both sides of this position. As shown by the blue curve in Fig.\ \ref{fig:3}(a), the near-field pulse duration varies from 35 fs to 118 fs, first decreasing and then increasing. Near the initial position, the compressor grating separation was adjusted more finely to resolve the detailed variation in pulse duration, as shown by the blue curve in the inset.

For each compressor setting, five measurements were performed to assess the repeatability of the retrieved far-field pulse duration. The orange curve in Fig.\ \ref{fig:3}(a) shows the mean of the repeated measurements, and the error bars denote the standard deviation. Data points measured near the initial position are generally more reliable.

The results show that the pulse durations measured before the focusing optics and those retrieved in the focal region follow the same overall trend and are comparable in magnitude. The remaining discrepancy can be attributed to transmissive elements in the diagnostic beam path, spatiotemporal coupling introduced by the focusing optics, and intensity-dependent nonlinear effects near focus \cite{Steiniger2024}. These observations further motivate direct in situ measurement of the pulse duration at the actual laser focus.

Figures \ref{fig:3}(b1)--(d1) show diffraction images recorded under several representative compressor settings. In Figs.\ \ref{fig:3}(b2)--(d2), the orange curves are lineouts obtained by integrating the two-dimensional diffraction images, while the blue curves are near-field pulse-duration traces measured with an autocorrelator. The pulse duration is determined by fitting the autocorrelation trace with either a $\mathrm{sech}^2$ or a Gaussian temporal profile. Using the approximately linear relation between grating length and pulse duration in the calibration curve, the horizontal axis of the integrated curves is converted from pixels to time. This conversion allows direct comparison with the autocorrelator results and provides a check on the measurement accuracy.

Beyond the agreement in FWHM, the overall profile shape is also consistent between the single-shot retrieval and the autocorrelator reference, including both the central lobe and the weak pedestal. The residual asymmetry seen in several traces is attributed to uncompensated higher-order dispersion and weak spatiotemporal coupling near focus. These effects are expected to become more pronounced in larger-aperture systems.

The remaining differences between the two sets of results can be understood in terms of the factors discussed above. In the present small-aperture laser system, these effects are relatively weak, and the near-field measurement provides a reliable reference for comparison. In large-aperture PW-class laser systems, however, the discrepancy between near-field and far-field pulse durations can be much larger.

\subsection{Scan reconstruction cross-check and aperture effect}

\begin{figure}[!htbp]
\centering
\includegraphics[width=10cm]{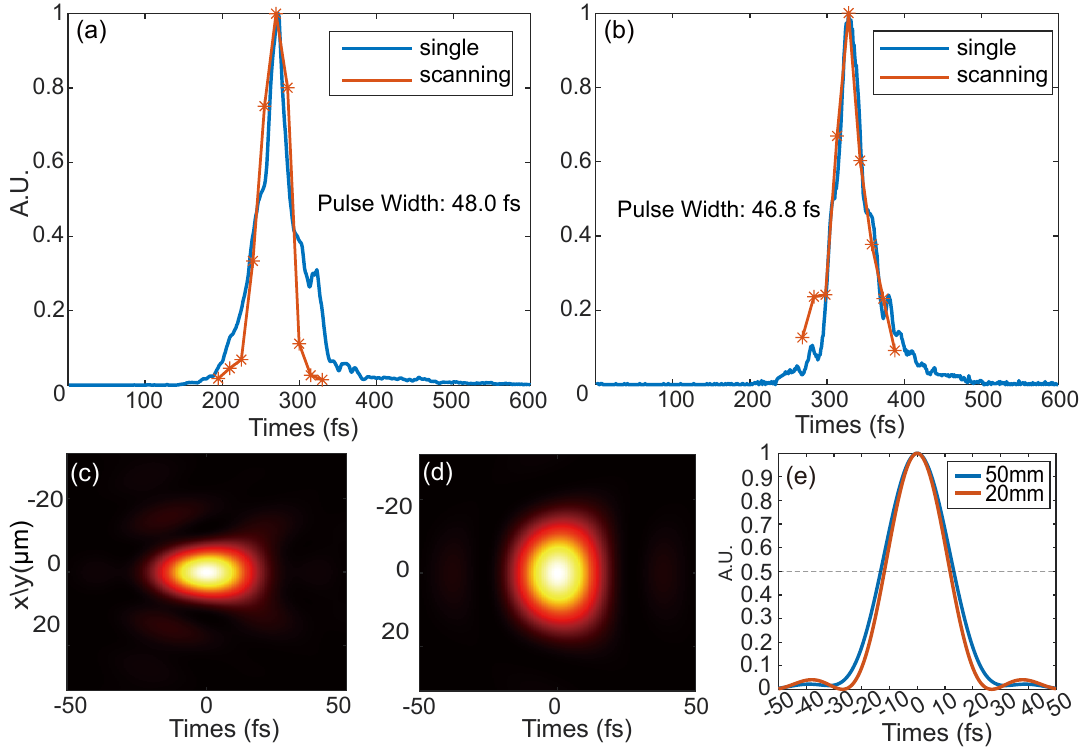}
\caption{Comparison of far-field temporal profiles obtained by scanning reconstruction and single-shot measurement for: (a) 50 mm-aperture incident laser beam. (b) 20 mm-aperture incident laser beam. (c)--(d) Simulated $x/t$ distributions after beam reduction with 50 mm and 20 mm apertures, respectively. (e) Simulated far-field temporal traces at $x=0$ for the two aperture sizes.}
\label{fig:4}
\end{figure}

To further validate the method, we performed a scan measurement analogous to scanning autocorrelation. The delay line in the signal-pump arm was adjusted continuously, so that the integrated diffraction image corresponded to the laser intensity at a given temporal slice, allowing the full temporal profile of the signal pump to be scanned. Because the minimum step size of the delay line was 15 fs, a compressor setting that yielded a pulse duration of about 45 fs was chosen for this measurement. As shown in Fig.\ \ref{fig:4}(a), the delay was scanned from $-45$ fs to $45$ fs. The orange curve shows the scanning result, while the blue curve shows the single-shot temporal trace. The close agreement between the two supports the reliability of the single-shot far-field pulse-duration diagnostic.

The scan reconstruction has a finite temporal sampling interval, 15 fs in the present setup, so narrow features on a comparable timescale can be smoothed by discrete sampling and interpolation. The scanning result is therefore used here mainly as an independent check of the overall envelope shape and width, whereas the single-shot method remains the only approach that provides true one-shot acquisition in low-repetition-rate facilities.

For ultrashort-pulse lasers, aberrations in the measurement optics must be controlled carefully. Before entering the diagnostic branch, the laser passes through a beam-reducing telescope, which may introduce spherical aberration. It is therefore important to verify that this effect does not significantly bias the measurement. To assess its influence experimentally, the input beam diameter was reduced from 50 mm to 20 mm. As shown in Fig.\ \ref{fig:4}(b), the retrieved far-field pulse duration changed only slightly, within 2 fs.

A simple numerical simulation was also carried out. A Fourier-transform-limited pulse was assumed, and the beam-reducing telescope was represented by an equivalent phase-screen model of a thick lens to evaluate its influence on the far-field pulse duration. Figures \ref{fig:4}(c)--(d) show the $x/t$ distributions after beam reduction, for 50 mm and 20 mm apertures, respectively, followed by focusing to the focal point. Figure \ref{fig:4}(e) shows the far-field temporal trace at $x=0$. The simulation indicates that reducing the beam aperture shortens the pulse by only about 3 fs, implying that spherical aberration in the measurement system does not significantly affect the retrieved result. These numerical results are consistent with the experimental observations in Figs.\ \ref{fig:4}(a)--(b).

The aperture-variation test and the numerical phase-screen analysis together provide a practical upper bound on the systematic bias introduced by the diagnostic branch, on the order of a few femtoseconds in the present platform. This error is smaller than the pulse-duration variation caused by compressor detuning, supporting the use of the present system for routine focal-region optimization.

\subsection{Intensity range and robustness}

\begin{figure}
    \centering
    \includegraphics[width=10cm]{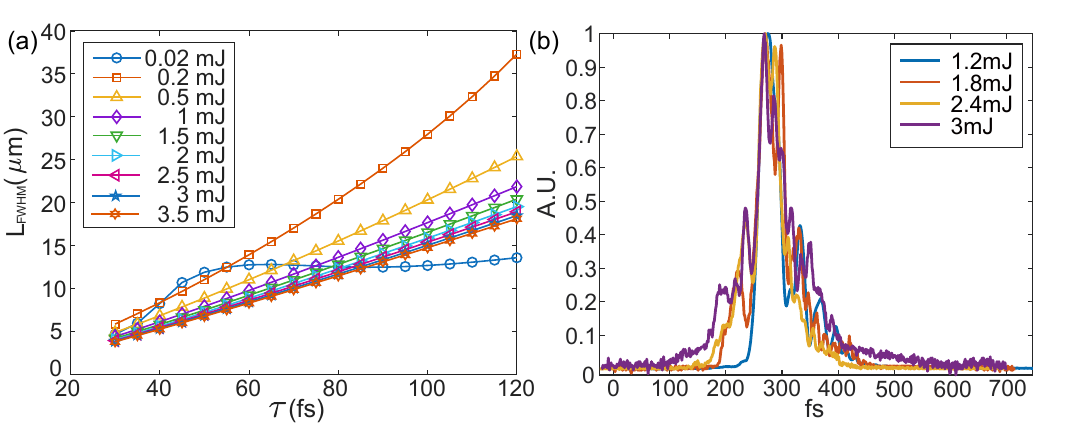}
    \caption{(a) Pulse-duration calibration curves calculated at multiple pulse energies. (b) Experimental results obtained by varying the pulse energy at a fixed compressor setting, showing the retrieved far-field temporal profiles.}
    \label{fig:5}
\end{figure}

In potential applications, the usable intensity range and robustness of the method are important. Using the model described above, we calculated a series of calibration curves $L(\tau)$ for pulse energies from 20 $\mu\mathrm{J}$ to 3.5 mJ, as shown in Fig.\ \ref{fig:5}(a). When the pulse energy exceeds 0.1 mJ, the calibration curve shows a stable linear relation. Small variations in pulse energy also have little effect on the slope of the curve.

For experimental validation, the compressor setting corresponding to optimal dispersion compensation was selected, and the output pulse energy was varied from 1.2 mJ to 3 mJ. Figure \ref{fig:5}(b) shows diffraction images recorded under different conditions, with the horizontal axis converted to time using the corresponding calibration curves. For the same compressor setting, the differences among the retrieved temporal profiles at different energies are small. At the highest energies, small spikes appear near the edges of the retrieved profiles. These features are attributed to imperfect realignment of the pump and probe foci after the energy increase, most likely due to thermal drift and other environmental perturbations. Importantly, these artifacts do not affect the validation of the single-shot measurement.

With the present experimental configuration, the method can stably retrieve the pulse duration in the focal region at intensities of $10^{16-17}\mathrm{W/cm^2}$, and it remains robust against shot-to-shot energy fluctuations in ultra-intense laser facilities. Potential limitations arise when the focal intensity approaches $10^{18}\mathrm{W/cm^2}$ or higher. Under these conditions, the gas medium tends toward near-complete ionization, which weakens the nonlinear dependence of electron density on laser intensity, and the plasma no longer remains in a regime mainly determined by the spatial intensity modulation of the laser \cite{keldysh_2014,PhysRevA.88.063421,PhysRevLett.63.2212,PhysRevLett.90.053002}. In addition, the high electron density introduces strong defocusing and phase distortion \cite{couairon_femtosecond_2007}. These effects reduce the modulation depth of the plasma grating and can substantially distort the rising edge of the signal pump pulse, thereby invalidating the underlying measurement principle. Under the current experimental constraints, the validity of the method at $10^{18}\mathrm{W/cm^2}$ has not yet been verified.

Several strategies may help extend the method to higher intensities. Replacing the medium with inert gases such as helium can raise the ionization threshold and may allow pulse-duration measurements at intensities approaching $10^{18}\mathrm{W/cm^2}$ \cite{Augst1991}. The calibration curve $L(\tau)$ can also be extended to a multi-dimensional form that includes both peak intensity and gas species. By introducing additional measurable plasma characteristics as constraints, it should be possible to maintain the accuracy of pulse-duration retrieval in higher-intensity conditions.

\section{Conclusion}

In summary, we propose a single-shot method based on a plasma grating for direct measurement of the far-field pulse duration. Experiments on a Ti:sapphire laser system demonstrate far-field pulse-duration measurements from 35 to 130 fs, and the single-shot results are validated by comparison with near-field autocorrelator measurements and far-field scanning autocorrelation. The method is also shown to operate at peak intensities up to $10^{16}{\rm W/cm^2}$. It is largely independent of the laser central wavelength, suppresses background noise, and requires only a simple experimental configuration. The demonstrated range is limited by the laser parameters available in the present system, but in principle the method can be extended to 15--300 fs and to higher peak intensities. The influence of aberrations in the measurement system, including chromatic and spherical aberrations, is further evaluated and found to be negligible compared with the intrinsic pulse properties before the beam enters the diagnostic branch (within $<2$ fs). Future work will focus on extending the accessible peak intensity, reducing the loss of accuracy caused by saturation in the background gas, and optimizing the optical configuration to broaden the accessible range of far-field pulse durations.

\appendix

\renewcommand{\thesection}{\Alph{section}}
\renewcommand{\theequation}{\thesection\arabic{equation}}
\renewcommand{\theHequation}{\thesection\arabic{equation}}
\renewcommand{\theequation}{\thesection\arabic{equation}}
\renewcommand{\theHequation}{\thesection\arabic{equation}}
\newcommand{\appsection}[1]{%
  \refstepcounter{section}%
  \setcounter{equation}{0}%
  \setcounter{equation}{0}%
  \section*{APPENDIX \thesection: #1}%
}

\appsection{MO-PPT ionization model}
For a molecule with fixed orientation $R$ (Euler angles) relative to a linearly polarized laser field with envelope $F(t)$ and carrier frequency $\omega$, the MO-PPT (cycle-averaged) instantaneous ionization rate is written as~\cite{Zhao2016PRA,Popruzhenko2008PRL}
\begin{equation}
\begin{aligned}
w_{\mathrm{MO\text{-}PPT}}!\left(F(t),\omega;R\right)
&=\sum_{m'} \left|B_{m'}(R)\right|^{2},
w^{(m')}{\mathrm{PPT}}!\left(F(t),\omega\right),
\end{aligned}
\end{equation}
where the molecular-structure factor is~\cite{Tong2002PRA,Zhao2016PRA}
\begin{equation}
\begin{aligned}
B{m'}(R)&=\sum_{l,m} C_{lm},D^{,l}{m',m}(R),Q{l m'},\
Q_{l m'}&=(-1)^{(m'+|m'|)/2}\sqrt{\frac{(2l+1)(l+|m'|)!}{2(l-|m'|)!}} .
\end{aligned}
\end{equation}
Here $C_{lm}$ are the asymptotic molecular-orbital coefficients (HOMO), and $D^{,l}_{m',m}(R)$ is the Wigner rotation matrix from the molecular frame to the laboratory frame.

The PPT channel rate is commonly expressed in the form~\cite{PPT1966,PPT1967,Popruzhenko2008PRL}
\begin{equation}
\begin{aligned}
w^{(m')}{\mathrm{PPT}}(F,\omega)
&=\mathcal{A}{m'}(\gamma),
\left(\frac{2\kappa^{3}}{F}\right)^{\nu(m')}\times \exp!\left[-\frac{2\kappa^{3}}{3F},g(\gamma)\right],
\end{aligned}
\end{equation}
with
\begin{equation}
\begin{aligned}
\kappa&=\sqrt{2I_p},\qquad
\gamma=\frac{\omega\kappa}{F}, \
g(\gamma)&=\frac{3}{2\gamma}\left[\left(1+\frac{1}{2\gamma^{2}}\right)\mathrm{asinh},\gamma
-\frac{\sqrt{1+\gamma^{2}}}{2\gamma}\right].
\end{aligned}
\end{equation}
$I_p$ is the ionization potential, $\gamma$ is the Keldysh parameter, and $\nu(m')$ and $\mathcal{A}_{m'}(\gamma)$ (nonadiabatic/channel prefactor) are given explicitly in Refs.~\cite{Zhao2016PRA,Popruzhenko2008PRL}.

For a femtosecond pulse, the ionization probability is obtained by integrating the instantaneous rate,
\begin{equation}
P(R)=1-\exp!\left[-\int_{-\infty}^{+\infty} w_{\mathrm{MO\text{-}PPT}}(F(t),\omega;R),dt\right],
\end{equation}
and for an isotropic (randomly oriented) ensemble,
\begin{equation}
P_{\mathrm{ave}}=\frac{1}{4\pi}\int P(R),d\Omega .
\end{equation}

\appsection{Reconstruction of Plasma Grating Length}

The $+1$ diffraction order is imaged onto a CCD with a single lens. To account for geometric projection, the coordinate measured along the propagation direction of the $+1$ order is mapped back to object space, since the $z$-axis and the diffraction direction form an angle $\theta$ in the $x$--$z$ plane.

The recorded two-dimensional image is integrated perpendicular to the stripe direction to obtain a normalized one-dimensional axial intensity profile. To improve accuracy, the broadening introduced by the system point-spread function (PSF) and residual aberrations is removed from the raw envelope. With magnification $M$ and pixel size $p$, the corrected grating length $L$ is given by
\begin{equation}
L = \frac{p}{M \sin \theta} Q_{\mathrm{corr}},
\end{equation}
where $Q_{\mathrm{corr}}$ denotes the corrected pixel count of the axial envelope.

To ensure that the measured profile $I_{+1}(z)$ faithfully represents the first-order electron-density component of the plasma grating, the system is operated strictly in the small-signal Bragg-diffraction regime. Experimental verification shows that $I_{+1}$ scales linearly with the incident probe energy, so the envelope shape and the extracted $L$ remain independent of probe intensity.

\appsection{Numerical calibration of grating length versus pulse duration}

We use a one-dimensional ionization model to generate a numerical calibration set $\left[\left(\tau_i,L_i\right)\right]_{i=1}^N$ for retrieving the pulse duration $\tau$. The detailed formulation of the ionization model is given in Appendix A. To keep the calibration consistent with the measurement, the physical conditions used in the numerical calculation are matched to those in the experiment, including the gas medium, the properties of the signal pump, and the temporal structure of the pump pulses.

The numerical calibration proceeds as follows. Two pump beams interfere in the gas and produce an intensity modulation. Strong-field ionization then writes a free-electron-density modulation in the focal region, whose evolution is described by Eq.(3). When the plasma-induced refractive-index modulation is dominant and the readout remains in the weak-scattering (small-signal) regime, the $+1$ Bragg-diffraction intensity is approximately proportional to the squared magnitude of the first-order electron-density component, $I_{+1}\left(z\right)\propto\left|N_{e,1}\left(z\right)\right|^2$. The numerical calibration therefore follows the same construction as the experiment: the axial grating envelope is obtained and its FWHM is extracted, yielding the calibration length corresponding to a given pulse duration.

\appsection{Impact of pump-energy mismatch}
  \setcounter{figure}{0}
  \renewcommand{\thefigure}{D\arabic{figure}}
  \begin{figure}
        \centering
        \includegraphics[width=7cm]{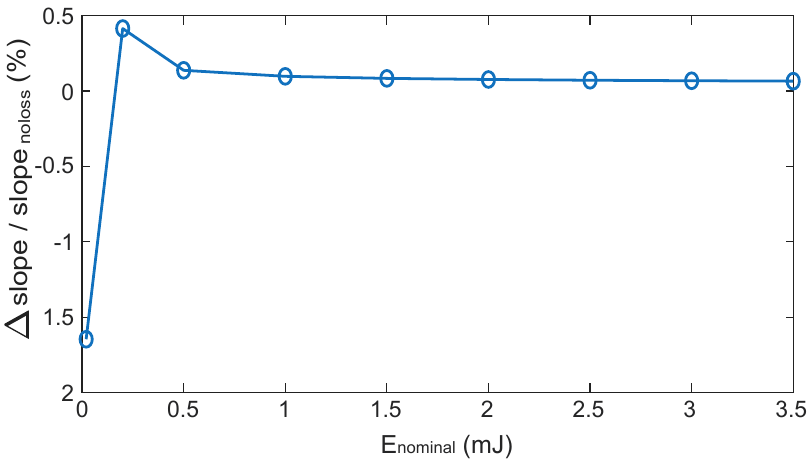}
        \caption{Relative deviation of the linear slope of the calibration curve from the ideal case when the pulse-energy mismatch between the reference pump and the signal pump is 1\%.}
        \label{fig:6}
  \end{figure}
As shown in Fig.\ref{fig:6}, the probe pulse is generated in the experiment by frequency doubling the reference pump, with a conversion efficiency of about 1\%. As a result, the signal-pump energy differs from the reference-pump energy by approximately 1\%. Under this condition, we compare the slope of the pulse-duration calibration curve with that of the ideal case at different energy levels. Within the target operating range, the relative slope deviation remains below 0.1\%, indicating that this energy mismatch has a negligible effect on the measurement accuracy and is smaller than the uncertainty associated with shot-to-shot pulse-duration fluctuations.

\begin{backmatter}

\bmsection{Funding}
National Key Research and Development Program of China (2022YFB3606305); National Natural Science Foundation of China (Grant No. 12475248); Natural Science Foundation of Guangdong (Grant No. 2025A1515012853); Natural Science Foundation of Top Talent of SZTU (Grant No. GDRC202310); Shenzhen Key Laboratory of Ultraintense Laser and Advanced Material Technology(Grant No. ZDSYS20200811143600001).

\bmsection{Acknowledgments}
The author, Jimin Wang, would like to express his sincere gratitude to Zitao Wang, Renhu Liu, and Zhilong Han for their invaluable assistance during the experiments, simulations, and manuscript preparation. The author also acknowledges the use of ChatGPT for linguistic refinement and LaTeX code generation in the preparation of this manuscript.

\bmsection{Disclosures}
The authors declare no conflicts of interest.

\bmsection{Data availability}
Data underlying the results presented in this paper are not publicly available at this time but may be obtained from the authors upon reasonable request.

\end{backmatter}

\bibliography{holography}

\begin{thebibliography}{10}
\newcommand{\enquote}[1]{``#1''}

\bibitem{Li2025}
Y.~Li, L.~Chen, M.~Chen, \emph{et~al.}, \enquote{High-intensity lasers and research activities in {China},} {\protect\JournalTitle{High Power Laser Sci. Eng.}} \textbf{13}, e12 (2025).

\bibitem{Palaniyappan2012}
S.~Palaniyappan, B.~M. Hegelich, H.-C. Wu, \emph{et~al.}, \enquote{Dynamics of relativistic transparency and optical shuttering in expanding overdense plasmas,} {\protect\JournalTitle{Nature Phys.}} \textbf{8}, 763 (2012).

\bibitem{Chen2015}
H.~Chen, F.~Fiuza, A.~Link, \emph{et~al.}, \enquote{Scaling the yield of laser-driven electron-positron jets to laboratory astrophysical applications,} {\protect\JournalTitle{Phys. Rev. Lett.}} \textbf{114}, 215001 (2015).

\bibitem{Macchi2013}
A.~Macchi, M.~Borghesi, and M.~Passoni, \enquote{Ion acceleration by superintense laser-plasma interaction,} {\protect\JournalTitle{Rev. Mod. Phys.}} \textbf{85}, 751 (2013).

\bibitem{Teubner2009}
U.~Teubner and P.~Gibbon, \enquote{High-order harmonics from laser-irradiated plasma surfaces,} {\protect\JournalTitle{Rev. Mod. Phys.}} \textbf{81}, 445 (2009).

\bibitem{Thaury2010}
C.~Thaury and F.~Qu{\'e}r{\'e}, \enquote{High-order harmonic and attosecond pulse generation on plasma mirrors: basic mechanisms,} {\protect\JournalTitle{J. Phys. B: At. Mol. Opt. Phys.}} \textbf{43}, 213001 (2010).

\bibitem{Zewail1988}
A.~Zewail, \enquote{Laser femtochemistry,} {\protect\JournalTitle{Science}} \textbf{242}, 1645 (1988).

\bibitem{Trebino1997}
R.~Trebino, K.~DeLong, D.~Fittinghoff, \emph{et~al.}, \enquote{Measuring ultrashort laser pulses in the time-frequency domain using frequency-resolved optical gating,} {\protect\JournalTitle{Rev. Sci. Instrum.}} \textbf{68}, 3277 (1997).

\bibitem{Shea2001}
P.~{O'Shea}, M.~Kimmel, X.~Gu, and R.~Trebino, \enquote{Highly simplified device for ultrashort-pulse measurement,} {\protect\JournalTitle{Opt. Lett.}} \textbf{26}, 932 (2001).

\bibitem{Bates2010}
P.~Bates, O.~Chalus, and J.Biegert, \enquote{Ultrashort pulse characterization in the mid-infrared,} {\protect\JournalTitle{Opt. Lett.}} \textbf{35}, 1377 (2010).

\bibitem{Radunsky2007}
A.~Radunsky, I.~Walmsley, S.-P. Gorza, and P.~Wasylczyk, \enquote{Compact spectral shearing interferometer for ultrashort pulse characterization,} {\protect\JournalTitle{Opt. Lett.}} \textbf{32}, 181 (2007).

\bibitem{Trisorio2012}
A.~Trisorio, S.~Grabielle, M.~Divall, \emph{et~al.}, \enquote{Self-referenced spectral interferometry for ultrashort infrared pulse characterization,} {\protect\JournalTitle{Opt. Lett.}} \textbf{37}, 2892 (2012).

\bibitem{Miranda2012}
M.~Miranda, T.~Fordell, C.~Arnold, \emph{et~al.}, \enquote{Simultaneous compression and characterization of ultrashort laser pulses using chirped mirrors and glass wedges,} {\protect\JournalTitle{Opt. Express}} \textbf{20}, 688 (2012).

\bibitem{Oksenhendler2010}
T.~Oksenhendler, S.~Coudreau, N.~Forget, \emph{et~al.}, \enquote{Self-referenced spectral interferometry,} {\protect\JournalTitle{Appl. Phys. B}} \textbf{99}, 7 (2010).

\bibitem{Iaconis1998}
C.~Iaconis and I.~Walmsley, \enquote{Spectral phase interferometry for direct electric-field reconstruction of ultrashort optical pulses,} {\protect\JournalTitle{Opt. Lett.}} \textbf{23}, 792 (1998).

\bibitem{Loriot2013}
V.~Loriot, G.~Gitzinger, and N.~Forget, \enquote{Self-referenced characterization of femtosecond laser pulses by chirp scan,} {\protect\JournalTitle{Opt. Express}} \textbf{21}, 24879 (2013).

\bibitem{Strickland1985}
D.~Strickland and G.~Mourou, \enquote{Compression of amplified chirped optical pulses,} {\protect\JournalTitle{Opt. Commun.}} \textbf{55}, 447 (1985).

\bibitem{Long2022}
T.-Y. Long, W.~Li, H.-T. Xu, and X.~Wang, \enquote{Influence of spatiotemporal coupling distortion on evaluation of pulse-duration-charactrization and focused intensity of ultra-fast and ultra-intensity laser,} {\protect\JournalTitle{Acta Phys. Sin.}} \textbf{71}, 174204 (2022).

\bibitem{Giordmaine1967}
J.~Giordmaine, P.~Rentzepis, S.~Shapiro, and K.~Wecht, \enquote{Two-photon excitation of fluorescence by picosecond light pulses,} {\protect\JournalTitle{Appl. Phys. Lett.}} \textbf{11}, 216 (1967).

\bibitem{Bradley1974}
D.~Bradley and G.~New, \enquote{Ultrashort pulse measurement,} {\protect\JournalTitle{{Proc. IEEE}}} \textbf{62}, 313 (1974).

\bibitem{Horowitz1993}
M.~Horowitz, A.~Bekker, and B.~Fischer, \enquote{Broadband second-harmonic generation in ${Sr}_x{Ba}_{1-x} {Nb}_2{O}_6$ by spread spectrum phase matching with controllable domain gratings,} {\protect\JournalTitle{Appl. Phys. Lett.}} \textbf{62}, 2619 (1993).

\bibitem{Fischer2007}
R.~Fischer, D.~Neshev, S.~Saltiel, \emph{et~al.}, \enquote{Monitoring ultrashort pulses by transverse frequency doubling of counterpropagating pulses in random media,} {\protect\JournalTitle{Appl. Phys. Lett.}} \textbf{91}, 031104 (2007).

\bibitem{Saltiel2008}
S.~Saltiel, D.~Neshev, R.~Fischer, \emph{et~al.}, \enquote{Spatiotemporal toroidal waves from the transverse second-harmonic generation,} {\protect\JournalTitle{Opt. Lett.}} \textbf{33}, 527 (2008).

\bibitem{Yu2018}
J.~Yu, X.~Ouyang, L.~Zhou, \emph{et~al.}, \enquote{Experimental study on measuring pulse duration in the far field for high-energy petawatt lasers,} {\protect\JournalTitle{Appl. Opt.}} \textbf{57}, 3488 (2018).

\bibitem{Shlomo2025}
N.~Shlomo and E.~Frumker, \enquote{In situ characterization of laser-induced strong field ionization phenomena,} {\protect\JournalTitle{Light Sci. Appl.}} \textbf{14}, 166 (2025).

\bibitem{Malkin1999}
V.~Malkin, G.~Shvets, and N.~Fisch, \enquote{Fast compression of laser beams to highly overcritical powers,} {\protect\JournalTitle{Phys. Rev. Lett.}} \textbf{82}, 4448 (1999).

\bibitem{Thaury2007}
C.~Thaury, F.~Qu{\'e}r{\'e}, J.-P. Geindre, \emph{et~al.}, \enquote{Plasma mirrors for ultrahigh-intensity optics,} {\protect\JournalTitle{Nat. Phys.}} \textbf{3}, 424 (2007).

\bibitem{Edwards2023}
M.~Edwards, S.~Waczynski, E.~Rockafellow, \emph{et~al.}, \enquote{Control of intense light with avalanche-ionization plasma gratings,} {\protect\JournalTitle{Optica}} \textbf{10}, 1587 (2023).

\bibitem{Leblanc2017}
A.~Leblanc, A.~Denoeud, L.~Chopineau, \emph{et~al.}, \enquote{Plasma holograms for ultrahigh-intensity optics,} {\protect\JournalTitle{Nat. Phys.}} \textbf{13}, 440 (2017).

\bibitem{Edwards2024a}
M.~Edwards, N.~Fasano, V.~{Perez-Ramirez}, \emph{et~al.}, \enquote{Structured light from structured plasma: Manipulating extreme lasers with plasma optics,} {\protect\JournalTitle{CLEO 2024}} p. {ATh1H.4} (2024).

\bibitem{Wu2022}
Z.~Wu, Y.~Zuo, X.~Zeng, \emph{et~al.}, \enquote{Laser compression via fast-extending plasma gratings,} {\protect\JournalTitle{Matter Radiat. Extremes}} \textbf{7}, 064402 (2022).

\bibitem{Li2023}
Z.~Li, Y.~Zuo, X.~Zeng, \emph{et~al.}, \enquote{Ultraintense few-cycle infrared laser generation by fast-extending plasma grating,} {\protect\JournalTitle{Matter Radiat. Extremes}} \textbf{8}, 014401 (2023).

\bibitem{Wu2024}
Z.~Wu, X.~Zeng, Z.~Li, \emph{et~al.}, \enquote{Generation of subcycle isolated attosecond pulses by pumping ionizing gating,} {\protect\JournalTitle{Phys. Rev. Res.}} \textbf{6}, 013126 (2024).

\bibitem{Edwards2022}
M.~Edwards and P.~Michel, \enquote{Plasma transmission gratings for compression of high-intensity laser pulses,} {\protect\JournalTitle{Phys. Rev. Appl.}} \textbf{18}, 024026 (2022).

\bibitem{Edwards2024b}
M.~Edwards, N.~Fasano, A.~Giakas, \emph{et~al.}, \enquote{Greater than five-order-of-magnitude postcompression temporal contrast improvement with an ionization plasma grating,} {\protect\JournalTitle{Phys. Rev. Lett.}} \textbf{133}, 155101 (2024).

\bibitem{Suntsov2009}
S.~Suntsov, D.~Abdollahpour, D.~Papazoglou, and S.~Tzortzakis, \enquote{Femtosecond laser induced plasma diffraction gratings in air as photonic devices for high intensity laser applications,} {\protect\JournalTitle{Appl. Phys. Lett.}} \textbf{94}, 251104 (2009).

\bibitem{Shi2011}
L.~Shi, W.~Li, Y.~Wang, \emph{et~al.}, \enquote{Generation of high-density electrons based on plasma grating induced bragg diffraction in air,} {\protect\JournalTitle{Phys. Rev. Lett.}} \textbf{107}, 095004 (2011).

\bibitem{Durand2012}
M.~Durand, A.~Jarnac, Y.~Liu, \emph{et~al.}, \enquote{Dynamics of plasma gratings in atomic and molecular gases,} {\protect\JournalTitle{Phys. Rev. E}} \textbf{86}, 036405 (2012).

\bibitem{Zhang2021}
C.~Zhang, Z.~Nie, Y.~Wu, \emph{et~al.}, \enquote{Ionization induced plasma grating and its applications in strong-field ionization measurements,} {\protect\JournalTitle{Plasma Phys. Control. Fusion}} \textbf{63}, 095011 (2021).

\bibitem{Waczynski2024}
S.~Waczynsk, A.~Zingale, M.~Edwards, \emph{et~al.}, \enquote{High efficiency plasma gratings generated by laser-driven avalanche ionization,} {\protect\JournalTitle{CLEO 2024}} p. {SM4Q.5.} (2024).

\bibitem{yeh1994introduction}
P.~Yeh and W.~Moerner, \enquote{Introduction to photorefractive nonlinear optics,}  (1994).

\bibitem{boyd2008nonlinear}
R.~W. Boyd, A.~L. Gaeta, and E.~Giese, \enquote{Nonlinear optics,} in \emph{Springer Handbook of Atomic, Molecular, and Optical Physics,}  (Springer, 2008), pp. 1097--1110.

\bibitem{Wahlstrand2011}
J.~K. Wahlstrand and H.~M. Milchberg, \enquote{Effect of a plasma grating on pump-probe experiments near the ionization threshold in gases,}  (2011). ArXiv preprint.

\bibitem{Steiniger2024}
K.~Steiniger, F.~Dietrich, D.~Albach, \emph{et~al.}, \enquote{Distortions in focusing laser pulses due to spatio-temporal couplings: an analytic description,} {\protect\JournalTitle{High Power Laser Science and Engineering}} \textbf{12}, e25 (2024).

\bibitem{keldysh_2014}
S.~V~Popruzhenko, \enquote{Keldysh theory of strong field ionization: history, applications, difficulties and perspectives,} {\protect\JournalTitle{Journal of Physics B: Atomic, Molecular and Optical Physics}} \textbf{47}, 204001 (2014).

\bibitem{PhysRevA.88.063421}
E.~Yakaboylu, M.~Klaiber, H.~Bauke, \emph{et~al.}, \enquote{Relativistic features and time delay of laser-induced tunnel ionization,} {\protect\JournalTitle{Phys. Rev. A}} \textbf{88}, 063421 (2013).

\bibitem{PhysRevLett.63.2212}
S.~Augst, D.~Strickland, D.~D. Meyerhofer, \emph{et~al.}, \enquote{Tunneling ionization of noble gases in a high-intensity laser field,} {\protect\JournalTitle{Phys. Rev. Lett.}} \textbf{63}, 2212--2215 (1989).

\bibitem{PhysRevLett.90.053002}
A.~Maltsev and T.~Ditmire, \enquote{Above threshold ionization in tightly focused, strongly relativistic laser fields,} {\protect\JournalTitle{Phys. Rev. Lett.}} \textbf{90}, 053002 (2003).

\bibitem{couairon_femtosecond_2007}
A.~Couairon and A.~Mysyrowicz, \enquote{Femtosecond filamentation in transparent media,} {\protect\JournalTitle{Physics Reports}} \textbf{441}, 47--189 (2007).

\bibitem{Augst1991}
S.~Augst, D.~D. Meyerhofer, D.~Strickland, and S.~L. Chin, \enquote{Laser ionization of noble gases by coulomb-barrier suppression,} {\protect\JournalTitle{Phys. Rev. Lett.}} \textbf{67}, 1486 (1991).

\bibitem{Zhao2016PRA}
S.-F. Zhao, A.-T. Le, C.~Jin, \emph{et~al.}, \enquote{Analytical model for calibrating laser intensity in strong-field-ionization experiments,} {\protect\JournalTitle{Phys. Rev. A}} \textbf{93}, 023413 (2016).

\bibitem{Popruzhenko2008PRL}
S.~V. Popruzhenko, V.~D. Mur, V.~S. Popov, and D.~Bauer, \enquote{Strong field ionization rate for arbitrary laser frequencies,} {\protect\JournalTitle{Phys. Rev. Lett.}} \textbf{101}, 193003 (2008).

\bibitem{Tong2002PRA}
X.~M. Tong, Z.~X. Zhao, and C.~D. Lin, \enquote{Theory of molecular tunneling ionization,} {\protect\JournalTitle{Phys. Rev. A}} \textbf{66}, 033402 (2002).

\bibitem{PPT1966}
A.~M. Perelomov, V.~S. Popov, and M.~V. Terent'ev, \enquote{Ionization of atoms in an alternating electric field,} {\protect\JournalTitle{Sov. Phys. JETP}} \textbf{23}, 924 (1966).

\bibitem{PPT1967}
A.~M. Perelomov, V.~S. Popov, and M.~V. Terent'ev, \enquote{Ionization of atoms in an alternating electric field: Ii,} {\protect\JournalTitle{Sov. Phys. JETP}} \textbf{24}, 207 (1967).

\end{thebibliography}

\end{document}